\documentstyle[preprint,tighten,eqsecnum,epsf,aps]{revtex}
\input{psfig}

\def\beq{\begin{equation}}   \def\eeq{
\end{equation}}

\begin{document}
\title{Space-Time  Evolution of the Oscillator, Rapidly
Moving  in a Random Media } \author{
B. Blok\thanks{E-mail: blok@physics.technion.ac.il}
}
\address{Department of Physics, Technion -- Israel Institute of
Technology, Haifa 32000, Israel}
\maketitle

\thispagestyle{empty}

\begin{abstract} We study  the
quantum-mechanical evolution of the nonrelativistic oscillator, rapidly
moving
 in the media with the random vector fields.
 We calculate the evolution of the level probability distribution
 as a function of time, and obtain rapid level diffusion over the energy
levels. Our results imply a new mechanism of charmonium
dissociation in QCD media.

 \end{abstract}

 \setcounter{page}{1} \section{Introduction.}
\par The propagation of the charged particles, both quantum and classical,
through different types of media had long ago become an important
branch of theoretical physics. (See e.g. an excellent review
\cite{AS} for the discussion on the  present status of the
subject). However, much less is known about the propagation of
neutral bound systems and the wave packets  with the zero total
charge but nonzero higher multipole moments. This is especially
true for quantum systems. Recently there was a revived interest in
the subject.
 This is  due to numerous possible applications in
nuclear and particle physics, especially in the study of
exclusive processes and  the creation and the diagnostics of the
quark-gluon plasma \cite{FS}. Such systems can be also studied
in electrodynamics, where it is possible to observe the
propagation of $e^+e^-$ positronium through the random media.
\par Although a significant advance in the study of the propagation of
these objects was made recently, very little is known about  the
evolution of the fast wave
the packets and the dipoles in the random media, and it's influence on the
particle and
nuclear cross-sections, diffractive processes and energy losses.
\par  The aim of this paper is to study the  propagation
 of the highly energetic, but nonrelativistic (in it's c.m.
rest reference frame) electrically neutral dipoles and wave packets, with
nonzero dipole moment, and with the internal interactions, through the random
media. For simplicity, and
in order to have an almost explicit solution we shall consider the
simplest example of such kind $-$ a  harmonic oscillator moving
through the  media with the random vector fields. The latter can be any
 media, that
where there are  random vector fields, i.e. the fields transforming as
vectors
under the Lorentz group transformations. In nature these are electromagnetic
and
Yang-Mills fields. We shall call below any such media a vector media. The
examples of the vector random media include amorphous solid state media
\cite{LK}, usual electromagnetic plasma \cite{Landau},
 dense QCD media \cite{H} , and quark-gluon plasma \cite{GSY,CGS}.
It is less clear what happens in the nuclear media or nuclei,
where the fields are colorless and it is not obvious how they
transform under Lorentz transformations.
 \par In our paper we consider the
nonrelativistic
 oscillator, whose center of mass moves with
the relativistic speed through the random media, which is
characterized by the static random electric  fields.
In our simple model, the particles inside the dipole interact via
harmonic oscillator potential (no Coulomb interactions between them),
while the interaction with the external field is electromagnetic.
However we shall argue below that at least qualitatively our results also
hold for the interaction with the QCD colored fields, and for the general
law of interaction between the constituents of the dipole.
\par Our main result will be the development of the general formalism that
will allow us to determine the density matrix, expansion rate and
level diffusion for nonrelativistic harmonic oscillator moving
with high velocity in the random media. We shall derive the effective
action and the effective transition operator for the bound system
propagation in the random media. This effective action
 will be used to study two closely connected problems: the
propagation of the small initial size dipole/wave packet through
the random media and of the oscillator in the ground state through
the random media. We shall determine the energy level
diffusion rate of the small wave packet for small and big times,
as well as the probability that it occupies level n after time t.
We shall also determine the expansion rate of the ground state and
the probabilities that after time t the oscillator will stay in
the ground state or will move into the state with the energy level  N.
 \par Taking into
account  that the electric fields transform as the components of
the rank 2 tensor during the Lorentz transformation, while the
time as the component of the 4-vector we shall obtain for a
certain interval of times the superdiffusion expansion law
for the wave packet density radius: \beq
\rho^2=(A_0t^2+A_1t^3)/\gamma^2 \label{super} \eeq Here $ A_0\sim d-1/d$
describes the quantum mechanical expansion rate of the wave packet
and$A_1$ is the coefficient
depending on the properties of the media, $\gamma$ is the Lorentz
factor and,$\rho$ is the  radius of the dipole, and d is the inverse
squared radius of the initial state (for the ground state d=1). Thus,
random media may strongly
influence the cross-sections and energy losses. The physical reason for
"superdiffusion" is the extraction of the resonant (i.e. with
oscillator proper frequency) mode from the spectrum of the random
electric field fluctuations.
\par This superdiffusion effect is very similar to the one discussed
recently in refs. \cite{G1,G2} in the context of the random walk
of the classical particles in the time dependent random potential
in
 statistical mechanics.
\par Our results imply a new mechanism for charmonium dissociation in
vector media:it is excited to higher energy levels due to the
scattering on random electric fields, and then dissociates into $D\bar
D$ pairs \cite{Susskind,Quigg}.
\par  The calculation will be made
in the eikonal approximation, i.e. we shall assume that the center
of mass of the  dipole moves along the straight line with the
ultrarelativistic constant velocity $v_0\approx c$, where $c$ is
the speed of light. We shall neglect in this approximation both
the deviations of the center of mass from the straight line, and
the  velocity changes due to the interaction with the random
fields. The latter corrections will be taken into account
elsewhere. We shall assume that the size of our dipole is small
relative to the inhomogeneity scale of the media $\approx
1/\kappa$.
\par As it was mentioned above, all our calculations will be carried in
the simplified model of the dipole with the oscillator interaction
between the constituents and electromagnetic interaction with the
external field. We neglect the Coulomb interaction between
internal constituents of the dipole. We expect that our result
will not change qualitatively if we generalize the external field
to a color one, and take into account Coulomb interaction between
the constituents.
\par We shall see that the influence of the random media on the oscillator
has very transparent mechanical analogue in the classical problem
of the excitation of the oscillator by the external resonance
force \cite{LLM}. The random media obviously has the role of this external
force. On the other hand previous investigations, using this
analogy, correspond to the study of the proper oscillations of the
oscillator.
\par Some of the aspects of the problem of the dipole propagation
 were considered
before \cite{FS}, \cite{FFS},\cite{Kopeliovich}, but the latter authors
did not take into account the influence of the media on the propagation of
the dipole and it's expansion, and the change of it's quantum
state during the propagation. They only considered the interaction
free dipole, neglecting the influence of the media on the dipole
structure and the internal interactions. Although as we shall see
the internal interactions do not lead to new qualitative effects
(as it was noted in ref. \cite{FFS}), the interaction with the
media does play a major role if time is not too small.
 The only previous attempt to
take into account the influence of the media on the
 propagation of the
oscillator in the media was made in ref. \cite{FS,Kopeliovich},
but in the different context. The latter authors considered not
vector, but scalar media, in particular nuclei, and parameterized
it's influence by taking into account absorption.
\par The problem of harmonic oscillator propagation
naturally arises in the context of the propagation of the bound
systems like charmonium through the  quark-gluon plasma.
\par Note that there are strictly speaking two different problems that
can be discussed in the connection with the oscillator: one is the
propagation of the two-particle bound state-dipole through the
media, another is the propagation of the wave packet. In the
approximation used in this paper we shall see that technically
this is the same problem. The quantum dynamics of the bound state
can be reduced to the quantum dynamics of the wave packet. So
below, if not stated otherwise we shall not distinguish between
them. However, beyond the approximation of the homogeneous
electric field discussed below, these are two different problems,
with different Lagrangians.
\par The paper is organized as follows. In the second chapter
we shall develop the general formalism for the description of the
quantum dipole moving through the random media. First, we shall
derive the effective Lagrangian for the propagating oscillator in
this media, which we shall model  by the usual plasma.
We shall neglect the radiation.
The discussion of this important phenomena will be considered in a
separate paper.
 Next we shall
consider the simplified model, that leads to the same qualitative
results. We shall show that in the eikonal approximation it is
sufficient to consider the  behavior of the oscillator in the
time-dependent homogeneous electric field.  We shall calculate
using the functional integral method the effective evolution
operator and the probability  to find an oscillator at particular
levels as a function of time. In the third chapter we shall find
the averaged density matrix and study the expansion of the small
wave packets and see the influence of the media on the wave
packet expansion. In the fourth chapter we shall consider the
probability evolution for the ground state wave packets and apply
our results to the simplest model of charmonium. We summarize our
main results in conclusion. Some details of the calculations will be
discussed in Appendices.
\section{The  Model of the Oscillator Evolution.}
\subsection{The  Formulation of the Problem.}
\par The most widespread model of the random  media where the bound system
can propagate
 is  the conventional model of plasma  \cite{Landau}. Such model
can be used qualitatively to describe the amorphous material.
 The media is represented
by a random set of ions that generate electric potentials with the Debye
screening.
\beq V(r)=g\exp(-\kappa r)/r \label{potential} \eeq Here g is the
electric charge. If we neglect in this model three particle
correlations, the potential$-$potential correlation function in
such system is (\cite{Landau})
 \beq L(\vec r
-\vec r' )=<V(\vec r)V(\vec r')>=A^2\exp{(-\kappa\vert
\vec r-\vec
r'\vert )}/\vert \vec r -\vec r'\vert \label{correlation} \eeq
 A is the constant dependent on the properties of the media. We
will assume that the correlations are gaussian, i.e.that all
correlation functions can be expressed through the two-point correlation
function (\ref{correlation}). This is the usual assumption in the
statistical mechanics. More complicated systems with nongaussian
correlations are usually highly nontractable.
\par For the
QCD media or the quark-gluon plasma the most wide-spread model is
the model discussed in ref. \cite{DG}. This model is very similar
to the conventional model of plasma, discussed above, except that
instead of the conventional electric charges the particles have
the color charges $T^a$, that are the generators of the $SU(3)$
color group, while the interactions contain factors $g^2T^aT^b$
that are very similar to the charge products. If we take into
account only the binary interactions, the plasma becomes
color-diagonal, and we can neglect the color factors, that will
not make qualitative influence on our results.The potentials and
the correlation functions are the same as in eq. (\ref{correlation})
(except some numerical factors whose influence will be discussed in a
separate publication).
\par Note that in terms of the language of the statistical mechanics
for all cases (plasma, QCD media, quark-gluon plasma) our problem
is the well known problem of the description of the propagation of
the wave packet in the random gaussian media characterized by
given correlation functions  ( see Appendix A for a more detailed
discussion). Consequently we consider in this paper all types of
plasma as an electromagnetic plasma.
 \par Consider now the fast moving neutral system of two
particles (dipole) with a harmonic interaction between it's components
in the center of mass reference frame:
\beq
W(\vert \vec r-\vec r'\vert)=N\vert r-r'\vert^2
\label{interpotential}
\eeq
 The system is moving in a random media discussed above
 with a velocity
$v_0\approx c$, where $c$ is the speed of light.
 For this ultrarelativistic  speed we
can use the eikonal approximation. We  assume that the center of
mass moves along the straight line  with a constant velocity.
Radiation will be neglected.
\par According to our assumption the oscillator is nonrelativistic,
consequently we can separate the motion of two constituents of the
dipole into the motion of the center of mass and the relative
motion of the two particles ( see ref. \cite{LH} and Appendix B
for detailed discussion): \beq L=L_{\rm c.m.} +L_{\rm
n.r.}\label{sum1} \eeq In the eq. (\ref{sum1}) the first term
describes  the free motion of the center of mass with the speed
$v_0$, while the second is the Lagrangian of the relative
nonrelativistic  motion of the two components of the dipole in the
c.m. reference frame. The latter Lagrangian can be written  for
the case of the small dipole with the size $\ll 1/\kappa$, \beq
L_{\rm n.r.}=\frac{1}{2}M(\dot u^2-\omega_0^2 \vec u^2)- g(V(\vec
r_1,t)-V(\vec r_2,t)) \label{nr} \eeq Here V is the electric
potential in the oscillator c.m. frame, $\vec r_1, \vec r_2$ are
the coordinates of the dipole components in this frame, and $\vec
u=\vec r_1-\vec r_2$ is the relative coordinate of two
constituents of the dipole. The derivation of the Lagrangian
(\ref{nr}) is discussed in Appendix B.
\par For the small size dipole we may use
a Taylor
expansion: \beq V(\vec r_1,t)-V(\vec r_2,t)=(\vec r_1-\vec
r_2)\frac{\partial V(0,t)}{\partial \vec r} \label{tay} \eeq Here
V(0,t) is the electric potential in the c.m. reference frame at
the moment t (t here is the laboratory time) that corresponds to
the random potential in the point where the center of mass is
located, it's derivative is obviously the electric field in this
point. Thus for the small size dipole, with the size $\ll
1/\kappa$, we can neglect inhomogeneity of the field  and reduce
the problem of the dipole motion to the problem of the oscillator
in the homogeneous electric field \beq \vec E=-\frac{\partial
V(0,t)}{\partial \vec r} \label{E1} \eeq (measured in the
oscillator c.m. frame in the point corresponding to the oscillator
c.m.). In this approximation, and this is an approximation that we
will use in the present paper, the problem of the moving
oscillator is equivalent to the problem of the propagation of the wave
packet
in the electric field $\vec E$.
 For the larger dipole we must use
the full second term in the Lagrangian (\ref{nr}), taking into
account the inhomogeneity of the field.
\par The same separation can be made  for the nonrelativistic neutral wave
packet, whose center of mass moves with a relativistic speed. In
this case the nonrelativistic Lagrangian that describes the
space-time evolution of the dipole in it's rest frame can be
written as
 \beq L_{\rm
n.r.}=\frac{1}{2}M(\dot {\vec r}^2-\omega_0^2 \vec r^2)- gV(\vec r,t)
\label{nr2} \eeq Here $\vec r$ is the distance between two particles in
the rest frame of the dipole or the radius vector in the c.m. at rest
reference frame for a wave packet. V is the potential in the wave packet
c.m. reference frame.
 The Lagrangian
(\ref{nr}) in the approximation  (\ref{tay}) is evidently a particular
case of eq. (\ref{nr2}). \par   Let us  go to the rest frame of the
dipole. In the latter frame our problem is evidently a problem of
the nonrelativistic oscillator with the internal interaction given by eq.
(\ref{interpotential} and in the external electromagnetic field.
\par Electromagnetic field acting on the particles that built up the
oscillator can be easily found by making the Lorentz transformation
\cite{LLF}.
(Note that within the present approximation, since the relative motion is
nonrelativistic, the relative speeds of the two particles can be neglected to
perform the Lorentz
transformation according to c.m. speed $v_0$).
\par After the Lorentz transformation (see appendix B for details), the
laboratory frame 4-potential $(V,0,0,0)$ of the
electromagnetic fields becomes $(V_{\rm c.m.},0,0,A_3)$:
$$V_{\rm c.m.}=V\gamma\\,\,\,\,\,{\rm }
A_3=-V\gamma$$
The electromagnetic tensor  $\vec E,\vec B$ ($\vec B=0$ in the laboratory
reference frame) becomes:
 \begin{eqnarray}
E_{{\rm c.m.}1,2}&=& E_{1,2}\gamma\,\,\,\,\,\,{\rm  } ,{\rm }
E_{{\rm c.m.}3}= E_3\nonumber\\
B_{{\rm c.m.}1} &=&v_0E_2\gamma,{\rm } B_{{\rm c.m.}2}=
-v_0E_1\gamma , \,\,\, B_{{\rm c.m.}3}=0 \label{tr}
\end{eqnarray}
Here $\gamma $ is the Lorentz factor: \beq
\gamma=1/\sqrt{1-v_0^2/c^2} \label{gamma1} \eeq
 \par We see that the transverse electric field is greatly enhanced, while
the longitudinal field remains the same.
\par We also see that in addition to electric fields there arises a
magnetic field
equal to \beq
\vec B =\vec v_0\times\vec E
\label{B}
\eeq
\par Fortunately, this magnetic field can be neglected. First, the magnetic
field is of the same order as the electric one, but the Lorentz
force acting on the dipole is suppressed as $v/c<<1$, where $v$ is
 the relative speed of the dipole constituents, since we assume
dipole to be a nonrelativistic system. Second, the magnetic field
is the field with very short period and high frequency $\approx
\kappa\gamma$. If we consider the classical motion in such
periodic field, we shall come to the system of differential
equations for the particle moving in combined oscillator-Lorentz
force field, where we can use the  Floque theorem. The Floque
theorem \cite{Hill} says that the system of the ordinary
differential equations with periodic coefficients has the
solution:
$$x(t)=A(t)C(t)$$
where B is the periodic matrix function and A is the diagonal
matrix, with entries $\exp{\lambda t}$ , the coefficients
$\lambda$ are called the eigenvalues of the  problem. The matrixes
are $6\times 6$ and include both the 3 coordinates and its derivatives
in time. Due to the energy conservation law and the fact that the
classical energy of the oscillator is $\approx
(dx/dt)^2+\omega_0^2x^2$, it is clear that all  eigenvalues of the
problem must be imaginary and appear in pairs (since the solution
is real). Thus magnetic fields may contribute only to high
frequency small fluctuations along the classical trajectory
corresponding to the motion without the magnetic field.
Consequently, it's effect can be safely neglected. Clearly the
same is true for the quantum mechanical case, since  the
corresponding path integral is saturated by the classical
trajectories.
\par  We have seen that the fields that act on the oscillator in
it's rest frame  are time-dependent. In the eikonal approximation
the coordinate of the bound state mass center is $$z=v_0t=v_0T
\gamma,$$ where  T is the proper time in the oscillator frame, and
$\gamma$ is a Lorentz factor: \beq T=t/\gamma \label{propertime}
\eeq Using the time and frequency transformation and field
transformation laws we can write the correlation function of the
potentials in the oscillator rest frame. The correlation function
will become time-dependent:
 \beq <V(\vec r,T)V(\vec
r',T')>=A^2\gamma^2\exp (-\kappa q)/q \label{q} \eeq \beq
q=\sqrt{(x-x')^2+(y-y')^2+\gamma^2(z-z'+v_0(T-T'))^2}
 \label{q1}
\eeq
 In the latter equations x and y are the transverse coordinates
that are the same in both reference frames, while z-s are the
z-coordinates of the oscillator in the c.m. frame. Due to the
Lorentz contraction, we can neglect z dependence and therefore
assume $z=z'$
(this can be done in Quantum Mechanics, where we neglect radiation).
Recall that $T$ and $T'$ are the proper time of the
oscillator. Thus we see that our problem  of the moving oscillator
wave packet has been reduced to the problem of the
nonrelativistic oscillator at rest in the high frequency  external
electric field that is strong in the transverse plane, and it's
longitudinal component has the same strength as in the  media at
rest, with the random field correlator (\ref{q1}).
\par In the small oscillator limit discussed above, the electric field
can be considered as homogeneous, and we get the correlation
function\beq <V(T)V(T')>=A^2\gamma^2\exp (-\kappa q)/q \label{q} \eeq
 \beq
q=v_0\gamma\vert T-T'\vert\approx \gamma\vert T-T'\vert
 \label{q3}
\eeq Here and below except in the $\gamma$ factors we assume
$v_0=c$, and use the units where $c=1$.
\par We shall argue below that
the results obtained in the homogeneous field approximation hold
qualitatively even for the large scale oscillator.
\par The eq. (\ref{q1}) holds
also for the dipole, since in
this case we work from the
beginning in the homogeneous  field approximation
(\ref{E1}).
\subsection{The General Formalism.}
\par Now we have to consider  the problem of
the nonrelativistic oscillator in the random media, with the
correlation function of the scalar potentials defined by eq.
(\ref{q}). This
 problem is very similar to the problem of the propagation of
the electromagnetic wave in the random media characterized by the
gaussian distribution function (see e.g.  ref. \cite{ms}).
\par Our main goal is to find the level distribution of the oscillator at
the moment of time t if at the 0 moment he was in the state
characterized by the wave function $\Psi (\vec r,0)$, or
equivalently by a set of coefficients $a_n(0)$: \beq \Psi (\vec
r,0)=\Sigma a_n\Psi_n (\vec r) \label{expansion} \eeq Here $\Psi_n
$ are the wave functions of the stationary states of the pure
oscillator without the random fields: \beq \Psi_n
(x)=(M\omega_0/\pi)^{1/4}\sqrt{\frac{1}{2^nn!}}\exp (-M\omega_0
x^2/2) \label{u} \eeq (we shall choose the units  where the Planck
constant $\hbar=1$).
 The probability that at the time t the oscillator will be at
the level n
  is
given by \beq P(n,t)=<\vert a_n(t)\vert^2>\label{pr} \eeq The average is
over all field configurations. The simplest way to calculate the
latter average for the oscillator is to calculate the average
product of the Green functions: \beq F(\vec x_0,\vec y_0,\vec
x_1,\vec y_1,t)=<G(\vec x_0,\vec x_1,t)G^*(\vec y_0,\vec y_1,t)>
\label{product} \eeq Here G is the oscillator Green function in
the external field V, that describes time evolution of the
oscillator in the given realization of the field V. The average is
taken over all external fields V with a Gaussian weight. Once we
know the function $G(x_0,y_0;x_1,y_1,t)$, the time evolution of
the $a_n$ coefficients can be easily calculated: \beq \vert
a_n(t)\vert^2=\int
d^3x_0d^3y_0d^3x_1d^3y_1F(x_0,y_0,x_1,y_1,t)\Psi(x)\Psi(y)\Psi_n(x')\Psi_n(y')
\label{evolution} \eeq Here we assumed that at t=o the oscillator
was in the state described by the wave function $\Psi (x,0)$.
\par One possible way to carry the calculations is to use the functional
integrals method. The Green function of the 3D quantum oscillator
is well known and can be represented as
  (see e.g. ref. \cite{FH},\cite{Kleinert},\cite{Schulman} and
references therein): \beq G(x,y,t)=\int
dx(t)\exp(i\int^t_0\frac{1}{2}M(\dot x^2-\omega_0^2 x^2))ds
\label{greenosc1} \eeq Here the integration is over all paths with
\beq x(t)=x_1,x(0)=x_0 \label{no} \eeq Consider now the arbitrary
wave packet. The full functional integral we are interested in is
\begin{eqnarray}
F(x_0,y_0,x_1,y_1,t)&=&\int dx(t)\int d y(t)\int dV(x,t)
\exp(i(\frac{M}{2}((\dot x)^2-\omega_0^2x^2)-V(x)\nonumber\\
&-&(i\frac{M}{2}((\dot y)^2
-\omega_0^2y^2)-V(y)))\nonumber\\
&\times&\exp(-\int^t_0\int^t_0
dsds'd^3xd^3yV(x,s)K(x,s;y,s')V(y,s'))\nonumber\\
 \label{funct}
\end{eqnarray}
The operator K is just the inverse of the correlation function of
the potentials V, and is determined by the
properties of the media. The integration is over all paths with
\beq x(t)=x_1,y(t)=y_1;x(0)=x_0,y(0)=y_1 \label{boundary} \eeq
Note that in the eq. (\ref{funct}) the integration over the
oscillator paths is in Minkowski space, with the usual Minkowski
time t, and i before the action. However, in the part of the
functional integral connected with the integration over V there is
no i before the action, and the integration is purely statistical
one. This integration is analogous to Euclidean space functional
integral in the Quantum Field Theory, and it is evident that it has
the same nature as the coupling of oscillator to the finite
temperature thermal bath.
\par It is very hard to carry out the exact functional integration in
eq. (\ref{funct}) even though the corresponding functional
integrals over V and x,y are Gaussian. Indeed, integration over V
can be carried out. We can proceed in the standard way, as in the
theory of the random fields, by looking for extremum V trajectory,
and then substituting the result into the functional integral
(\ref{funct}) that will be now the integral only over x,y. In
order to carry the integration, we write the term V(x(t),t)) in
the action as $\int d^3x \delta (x-x(t))V(x,t)$. Varying the
action over V(x,t), we obtain: \beq i\delta (x-x(t)-i\delta
(y-y(t))=-\int^t_0 K(x,t;x't')V(x',t') \label{extr1} \eeq This is
the equation for the extremal trajectory. The latter equation
(except the factor i due to the fact that we carry the average in
the Minkowski space) looks like the potential created by two
opposite charges moving over the trajectories x(t) and y(t). The
equation can be inverted giving \beq V(x,t)=-i\int^t ds
(R(x,t;x(s),s)-R(x,t,y(s),s)/2 \label{V} \eeq Here R is the
correlation function (\ref{q1}), that is inverse to the operator
K. Substituting this expression for V in the functional integral
(\ref{funct}) we get \beq F(x,y,x',y',t)=\int d x(t)\int dy(t)
\exp{i(M/2)(\dot x^2-\dot y^2 -\omega_0 x^2+\omega_0
y^2)-S_A(x(t),y(t))} \label{intout} \eeq Here the effective action
A is
\begin{eqnarray}
S_A&=&\int^t_0 dt'\int^{t'}_0
ds(R(x(t'),t';x(s),s)+R(y(t'),t';y(s),s)\nonumber\\
&-&R(x(t'),t';y(s),s)-R(y(t');t';x(s),s))
\label{efaction}
\end{eqnarray}
Note that the above action looks quite similar to the famous
Feynman action for the polaron \cite{Feynman}. It describes
 two particles with the singular interaction potential,
that includes both interaction with it's own trajectory and the
interaction with the trajectory of the second particle-the
situation quite similar to the problem of electrostatic dipole.
\par It is possible to deal with the latter action using mean field
approximation or perturbation theory, however for fast particle
the problem can be significantly simplified.
\par In order to simplify the problem further, first note
that the main contribution into the action comes from the region
$t'\approx s$, due to the exponential Yukawa cutoff in G. In the
area $t'\approx s$, we can expand $$\vec x(t')-\vec x(s)\approx
-(d\vec x/dt')(t'-s). $$ Then the radius in the Yukawa correlation
function \ref{q1} will have the form \beq
q=\sqrt{(t'-s)^2\gamma^2(v_0^2+(d\vec x/dt')^2)} \label{mun} \eeq
Since the system is nonrelativistic, the second term in the brackets in eq.
(\ref{mun})
can be neglected, and the system reduces to the usual quantum mechanics, i.e.
we can neglect both the longitudinal and  the transverse
self-motion. Concerning the interaction with the trajectory of the
second particle, the contribution is clearly maximal when x(t) is
close to y(t), and in this case also we can neglect the transverse
motion. This means that for a relativistic particle in the first
approximation we can neglect the space dependence of the electric
field and assume that the system is in the homogeneous time
dependent random electric field. In fact it was clear from the
very beginning that we can take our electric field as homogeneous.
Indeed, if we look at the action of the electric field, we see
that the term due to the space inhomogeneity contributes into the
action as $\kappa^2E^2$, relative to $\kappa^2\gamma^2E^2$ due to
the time evolution, and thus can be neglected, even if the initial
wave packet is not of the small radius. We shall see below, that
the inhomogeneity of the electric field must be taken into account
only at the times of the order of the minimum of two scales:
$\gamma^{4/3}/(\omega_0^2\kappa)^{1/3},
\gamma^2/\sqrt{(d-1/d}\kappa\omega_0$,
when the longitudinal expansion of the wave packet becomes significant$-$of
the order of $1/\kappa $. Thus our results can be applied even
for the more general case when $1/\kappa\leq
 1/\omega_0$.
\par Concerning the small dipole, we deal with it in the approximation
of the homogeneous electric field from the beginning, the field being
equal to the field in the c.m. of the dipole.
\par Next, we need to deal with the Coulomb  singularity of the
Yukawa potential. We shall follow the line of ref.
\cite{Kleinert}.  The contribution of the singular part of the
interaction to the functional integral is zero: the probability of
such configuration is zero. Moreover, it is known that such
singularity really does not appear due to quantum fluctuations,
and it's appearance is due to the quasiclassical method of the
path integral calculation, since in the classical case there can
be fall on the center. Consequently, if we are interested in the
qualitative results, we can substitute the singular correlation
function with a nonsingular one, omitting the coulomb part and
dealing directly with the electric fields. As for the strength of
the electric field we can take the strength of the field it is
clear from eqs. (\ref{nr}),(\ref{E1}) that we have to take the field in
the reference frame where the center of mass  of the system is at
rest.
 \par  Thus we come  to the
conclusion that
both qualitatively in the leading eikonal approximation and
quantatively we can describe our model by an oscillator coupled
to the homogeneous time dependent random electric field with the
correlation function in the longitudinal direction given by \beq
<E_z(t)E_z(s)>=B^2\exp(-\kappa\gamma v_0\vert T-s\vert)
\label{corl} \eeq and  in the transverse plane with the
correlation function \beq
<E_i(t)E_j(s)>=(B^2\gamma^2)\exp(-\kappa\gamma v_0\vert
T-s\vert)\delta_{ij} \label{cor} \eeq Here T and s  denote the
proper time (in the moving oscillator frame of reference), and the
frequency is increased greatly by $\gamma$-a Lorentz factor. The
corresponding potential is $V(x)=\vec E\vec x$, and the
corresponding Lagrange density in the action  can be rewritten as
\beq \delta S_A=A^2\int dt E(t)(\partial^2+\kappa^2\gamma^2)E(t)dt
\label{action} \eeq
 Note that the
simplifications we did are very similar to the Feynman trick for
polaron, where he substituted the complicated nonlinear action of
a polaron by an action of the harmonic oscillator.
\subsection{The  Calculation of the Effective Action.}
\par Now we shall calculate the 4-point function F in the above model.
 In order to average over the field configurations we shall use the  action
\beq S=A^2\int ((\partial E)^2+\kappa^2\gamma^2E^2)dt
\label{action} \eeq The constant A is determined using the
relation with the correlation function of the oscillator: \beq
<E(t)E(t')>=B^2(1,\gamma^2)\exp{-(\vert t-t'\vert \kappa\gamma)}
\label{al} \eeq (the brackets depend on whether the field is in
longitudinal or transverse direction). Then for the longitudinal
field we have  : \beq E_L^2(0)=B_L^2=
1/(A_L^2(2\kappa\gamma)) \label{kl1} \eeq For the transverse field
we have\cite{Kleinert} : \beq E_T^2(0)=B_T^2\gamma^2=
1/(A_L^2(2\kappa\gamma)) \label{kl} \eeq Here both electric fields
$E_L$ and $E_T$ are the fields
 in the frame moving
with the velocity $v_0\approx c$ where the harmonic oscillator is
at rest , so that B is the $\sqrt E^2_L(0)$, where index L means
the laboratory system, and the time t here and below refers to the
frame moving with the oscillator. Hence for the longitudinal field
\beq A_L=1/(B_L\gamma^{(1/2)}\sqrt{2}) \label{rel1} \eeq For the
transverse field: \beq A_T=1/(B_T\gamma^{(3/2)}\sqrt{2})
\label{rel} \eeq
\par Let us now repeat the derivation of the 4-point function G
for the concrete example of the averaging weight $E^2$. The derivation is
slightly different from the previous chapter, since there we dealt with
potential V and varied over it, while for the final simplified model we
can deal directly with electric field and vary over it that simplifies the
problem.
\par Due to the obvious factorization we can start from the
one-dimensional problem. The whole calculation is carried in the rest
frame of the moving oscillator.
\par More explicitly, the functional integral we need to calculate is

\beq F(x,y,x',y',T)=\int dx(t) dy(t) dE(t)\exp{iS(x,,y,\vec
E)-S(E)} \label{aver} \eeq Here  S(x,y,E) is the action \beq
S=\int^T_0 dt\frac{M}{2}(\dot x^2-\omega_0^2 x^2-\dot
y^2+\omega_0^2 y^2)-\vec E\vec x+\vec E\vec y \label{first} \eeq
The statistical weight S(E) is given by eq. (\ref{action}). The
boundary conditions for x,y are evident. Note that the first term
in the functional integral (\ref{aver}) is multiplied by i while
the second is not. We first will carry the integration over the
random fields, and then consider the effective action for x and y.
The integral over the electric field E is Gaussian and can be
taken. We first need to vary the action S over the fields E and
then substitute the result into the action. Varying over E we get:
\beq i (-x_i(t)+y_i(t))=A^2(\partial^2-\kappa^2\gamma^2)E_i(t)
\label{vary} \eeq The latter equation can be immediately solved
using the Green function of the operator $(\partial^2-\kappa^2)$:
\beq (\partial^2-\kappa^2)\exp (-\vert\kappa
t-t'\vert)/(2\kappa)=\delta (t-t') \label{green} \eeq
Consequently, we write \beq
E_i(t)=-(i/(2A^2\kappa\gamma))\int^T_0\exp)(-\vert
t-s\vert)(x_i(s)-y_i(s))ds \label{equation} \eeq
\par  We now
substitute the above expression for E into the action in order to
get effective action for x and y. We have for the terms that
previously contained  the electric field \beq
S_2=-\int^T_0\int^T_0 dtds\exp(-\kappa\gamma\vert t-s\vert)
(x(t)-x(s) (y(t)-y(s))/(2A^2\kappa\gamma) \label{final} \eeq Note
that the internal integration over s in the action goes up to T
and not t, like in the Minkowski field case. The reason is that
the averaging over field configurations takes part only after the
concrete trajectory in x(t) is realized. Physically, we first
calculate functional integral for given E realization, and then
average over all E.
\par We now have the effective action $S_1+S_2$,
where \beq S_1=iM(\dot x^2-\dot y^2-\omega_0 x^2+\omega_0 y^2)/2
\label{ac1} \eeq The calculation of the functional integral
proceeds in the standard way. We seek the solution of the
classical equation of motion, giving the exponent, and then
calculate the preexponential factor.
\par Let us start from the equations of motion. We get
\beq iM(\frac{\partial^2 x}{\partial t^2}+\omega_0
x)=-\int^T_0\exp(-\kappa\gamma \vert t-s\vert (
x(s)-y(s))/(A^2\kappa\gamma ) \label{eq1} \eeq \beq
iM(\frac{\partial^2y}{\partial dt^2}+\omega_0 y)=-\int^T_0 \exp
(-\kappa\gamma\vert t-s\vert ) (x(s)-y(s)/(A^2\kappa\gamma )
\label{eq2} \eeq We see that the equations for x and y are
identical, thus \beq \frac{\partial^2 (x-y)}{dt^2}+\omega_0^2
(x-y)=0 \label{minus} \eeq The boundary conditions are: \beq
\delta_1=x(T)-y(T)=x_1-y_1;\delta_0= x(0)-y(0)=x_0-y_0
\label{bound} \eeq The corresponding solution is evidently \beq
x(t)-y(t)=(\delta_0 \sin (\omega_0
(T-t))+\delta_1\sin\omega_0t)/(\sin\omega T) \label{hom} \eeq Once
we know x$-$y, we can substitute it to eq. (\ref{eq2}) and get the
equation for y, which is just the equation for the oscillator
under the action of the external imaginary force.
\begin{eqnarray}
\frac{\partial^2 x}{dt^2}+\omega^2_0 x&=&i/((MA^2\kappa\gamma)\sin
(\omega_0 T)(\kappa^2\gamma^2 +\omega_0^2))\nonumber\\
&\times&(\exp (-\kappa\gamma t) (\omega_0\delta_1-\delta_0(\omega_0\cos
(\omega_0
T)+\kappa\gamma\sin (\omega_0 T))\nonumber\\
&+&\exp (-\kappa\gamma (T-t))(\delta_0\omega_0-\delta_1(\omega_0\cos
(\omega_0T)+\kappa\gamma\sin (\omega_0T))+\nonumber\\
&+&2\kappa\gamma (\delta_1\sin (\omega_0 t)-\delta_0 \sin (\omega_0
(t-T)))
\label{eq3}
\end{eqnarray}
Note that the external force that appears in the r.h.s. of the eq.
(\ref{eq3}) has three parts. Two of them are
exponentially suppressed and are different from zero only close to
t=0 or t=T. However, the third term is not suppressed, and,
moreover, is a resonance force applied to the harmonic oscillator
with a frequency $\omega_0$. This is of course what was to be
expected, since the random field distribution contains all
possible frequencies, including the resonant one. The solution for
x for given boundary conditions will be a sum of the solution
of the homogeneous equation plus the solution of the
nonhomogeneous one: \beq x(t)=\frac{1}{\omega_0}\int^t_0\sin
(\omega_0 (s-t)) u(s)ds+x_0\cos(\omega_0 t)+B\sin (\omega_0 t)
\label{u} \eeq Here u(s) is the r.h.s. of eq. (\ref{eq3}). The
coefficient B(T) is determined from the boundary conditions.
\par  The general solution of the nonhomogeneous equation is
a sum of five terms:
\beq
x(t)=x_a(t)+x_b(t)+x_c(t)+x_d(t)+x_e(t)
\label{sum}
\eeq
The first term is due to the resonant part of the external field and is
given by
\beq
x_a(t)=iL(\delta_1\kappa\gamma t\cos (\omega_0 t)-\delta_0\kappa\gamma t
\cos (\omega_0 (T-t)))/\omega_0
\label{res}
\eeq
The second part includes the terms that come from the integration of the
resonant part that are not
linearly enhanced as the term above:
\beq
x_b(t)=iL\frac{\kappa\gamma}{\omega_0^2}(\delta_0\sin (\omega_0 t)
\cos (\omega_0 T)-\delta_1\sin (\omega_0 t))
\label{nr5}
\eeq
Here
$$L=1/(A^2\kappa\gamma)M\sin (\omega_0 T)(\kappa^2\gamma^2+\omega_0^2))$$
The third part comes from the integration of the nonresonant
force, but is not exponentially suppressed, however it is
suppressed by additional power of $\kappa\gamma$ relative to the
previous term: \beq x_c(t)=(iL/(\kappa\gamma^2+\omega_0^2))
(\omega_0\delta_1-\delta_0(\omega_0\cos (\omega_0
T)+\kappa\gamma\sin (\omega_0T))(\omega\cos
(\omega_0t)-\kappa\gamma\sin (\omega_0 t)) \label{nr6} \eeq Finally
the 4th term is the exponentially suppressed part coming from the
nonresonant part of the external force:
\begin{eqnarray}
x_d(t)&=&(iL/(\kappa\gamma^2+\omega_0^2))(-\omega_0
(\omega_0\delta_1-\delta_0 (\omega_0\cos (\omega_0 T)+\kappa\gamma\sin
(\omega_0 T))\exp (-\kappa\gamma t)\nonumber\\
&+&(\delta_0\omega-\delta_1 (\omega\cos (\omega_0 T)+\kappa\gamma
\sin (\omega_0
T))\nonumber\\
&(&\exp (-\kappa\gamma T)(\kappa\gamma\sin (\omega_0
t)+\omega_0\cos (\omega t))-\omega\exp (-((T-t)))) \label{sup}
\end{eqnarray}
Finally there is a solution of the homogeneous equation that can be
rewritten as:
\begin{eqnarray}
x_e(t)&=&x_0\cos (\omega_0 t)+
(x_1-x_a(T)-x_b(T)-x_c(T)-x_d(T)\nonumber\\
&-&x_e(T) -x_0\cos (\omega_0 T))\sin (\omega_0 t)/\sin (\omega_0
T) \label{ho}
\end{eqnarray}
\par Note that  if we are interested in time scales
$$t_L\ge 1/\kappa$$
we can neglect the terms that are exponentially suppressed.  The
terms in $x_b$ have the same structure essentially as in $x_b(t)$
but are enhanced by at least $\kappa\gamma/\omega_0$. Hence we can
write for
 $t_L\ge 1/\kappa$
 \beq x(t)=x_a(t)+x_b(t)+x_e(t) \label{f} \eeq
 Here in
$x_e(t)$ we can in the expression for B(T) neglect all the terms
except coming from $x_a(T)+x_b(T)+x_0\cos (\omega_0 T)$. Then for
B(T) we have \beq B(T)=(x_1-x_0\cos (\omega_0 T)-iL\kappa\gamma (
T (\delta_1\cos(\omega_0 T)-\delta_0)+\sin (\omega T)(\delta_0
\cos (\omega_0 T)-\delta_1)/\omega_0)/\sin(\omega_0 T) \label{B}
\eeq We see that the classical trajectory of the oscillator in
random field corresponds to the sum of two terms: first, the usual
harmonic oscillations (due to the homogenic part of the solution),
second the linear expansion (although suppressed as $1/\gamma$) in
the imaginary direction .
\par We now calculate the action as the function of the boundary
conditions
using the solution (\ref{f}).  Due to the equations of
motion the action is just \beq i(\dot x (T)x_1-\dot
y(T)y_1-x_0\dot x(0)+\dot y(0) y_0), \label{ac} \eeq The other
terms in the action disappear due to the equations of motion. It
is easy to obtain:
\begin{eqnarray}
S&=&\frac{M\omega_0}{2}( (x_1^2-y_1^2-x_0^2-y^2_0) \cot (\omega_0
T))-2(x_0x_1-y_0y_1)/\sin(\omega_0 T)))\nonumber\\
&+&i(L\kappa\gamma /\sin (\omega_0
T))((\delta_1^2+\delta_0^2)(\omega T-\sin (2\omega
T)/2)+\nonumber\\&+&2\delta_0\delta_1(\sin (\omega_0 T)-\omega_0
T\cos (\omega_0 T)) \label{finaction}
\end{eqnarray}
This action  gives us the transition operator as a function of
boundary conditions. Note that it contains three types of terms:
1) real part; 2) Imaginary part with linearly enhanced terms due
to quantum fluctuations 3) Imaginary part with only trigonometric
terms. Note that we neglected for  all the terms that are
suppressed by $\exp (-T))$ and by the powers of $\gamma$. The
transition operator in the quasiclassical approximation will be \beq
F=\exp
(iS) \label{tr} \eeq
\par We need now to consider the preexponential factor, and normalize
to the evolution operator of the harmonic oscillator without electric
fields.
In order to calculate the preexponent we need to expand:
\beq
x(t)=x_{cl}(t)+\sum^\infty _0 c_n \sin (2\pi n t/T)
\label{ex}
\eeq
\beq
y(t)=y_{cl} (t) +\sum^\infty_0 b_n\sin (2\pi n t/T)
\label{y}
\eeq
Since the cross terms with $x_{cl}$ evidently cancel we need to consider
the integral in the action due to the sum over n.
It is equal to:
\begin{eqnarray}
P&=&\prod \exp (iM(b_n^2-c_n^2)((2\pi
n/T)^2+\omega_0^2)\nonumber\\
&-&\frac {1}{2A^2\kappa\gamma}
\sum (c_n-b_n)(c_m-b_m)\int^T_0\int ^T_0\sin (2\pi n s/T)\nonumber\\
&\times&\exp
(-\kappa\gamma\vert t-s\vert \sin (2\pi mt/T)dtds
\label{prod}
\end{eqnarray}
  We can now move to the integration over the variables
$u_n=b_n-c_n$ and $v_n=b_n+c_n$. Then we evidently get after the
integration over $v_n$ the product of delta-functions
$$\prod \delta (u_n ((2\pi n/T)^2+\omega_0^2))$$
Then we can evidently let $u_n=0$. The preexponential factor $P$
will be the same as the corresponding factor for the oscillator
without the electric field which is evidently: \beq
\prod^{n=\infty}_{n=0} \frac{1}{(2\pi n/T)^2+\omega_0^2)}=\sin
(\omega_0T)\label{ds}\eeq.
\par Now we can calculate the transition operator G, whose matrix elements
give the transition probabilities we are looking for. We obtain
\begin{eqnarray}
F(x_0,y_0;x_1,y_1;T)&=&\frac{1}{\sin (\omega_0 T)}\exp
((iM\omega/2)( \frac{1}{\sin (\omega_0 T)}
((x_0^2+x_1^2-y_0^2-y^2_1)\nonumber\\
&\cos&(\omega_0
T)-2(x_0x_1-y_0y_1)))\nonumber\\
&\times&\exp-(ML\kappa \gamma\frac{1}{\sin(\omega_0
T)}(\delta_1^2+\delta^2_0)
(\omega_0 T -\sin (2\omega_0 T)/2)\nonumber\\
&+&2\delta_0\delta_1(\sin (\omega_0 T)-\omega_0 T\cos (\omega_0
T))) \label{G}
\end{eqnarray}
Note that the first term is just the usual evolution operator for
the harmonic oscillator, while the second term contains the action
of the random field. The operator is written in terms of the proper time,
in the oscillator reference frame. In order to work in the laboratory
frame we need to substitute $T=t_L/\gamma$, where $t_L$ is the laboratory
time. We can  use the latter transition
operator to calculate transition amplitudes as the functions of
time.
\subsection{Transition Amplitudes.}
\par Once we know the transition operator (\ref{G}) we can calculate
the evolution of the initial distribution. If in the initial state t=0 the
oscillator is in the state $k$, the probability that after time t the
oscillator will be in the state k is
\begin{eqnarray}
\vert a_n^k(T)\vert^2&=&\int dx_1dx_0dy_1dy_0
H_n(x_1)H_n(y_1)H_k(x_0)H_k(y_0)\nonumber\\
&\exp& -(M\omega_0
/2)(x_0^2+y_0^2+x_1^2+y^2_1))F(x_0,y_0,x_1,y_1;T) \label{ampl}
\end{eqnarray}
Here $N_n$ is the normalization factor:
 \beq N_n= (M\omega_0/\pi)^{1/4}\sqrt{1/(2^nn!)}
\label{norm} \eeq

In order to carry the actual calculation we need to go to the variables
$s_0=x_0+y_0,s_1=x_1+y_1;\delta_0,\delta_1$ and carry out the
corresponding gaussian integral over the four variables. Note
that due to the cross-term between $\delta_0\delta_1$ in the transition
operator F we have
\beq
\vert <a>\vert^2\ne <\vert a\vert^2>
\label{ne}
\eeq
\par The easiest example to calculate is the transition
probability
$\vert a^0_0\vert^2$ for the oscillator entering the media in the
initial state, and then staying in the initial state.
 The corresponding gaussian integral is   \beq \vert
a^0_0 (T)\vert^2=1/\sqrt{H} \label{H} \eeq where H is given by
\beq H=1+2\omega_0 Dt_L+4D^2(\omega_0^2t_L^2-\gamma^2\sin^2
(\omega_0 t_l/\gamma )) \label{answer} \eeq Here \beq
D_L=2B_L^2\kappa/ (M\omega_0^2(\kappa^2\gamma^2 +\omega_0^2))
\label{D} \eeq for the longitudinal motion and \beq
D_T=2B_T^2\kappa\gamma^2/(M\omega_0^2) (\kappa^2\gamma^2
+\omega_0^2)) \label{D} \eeq for the transverse motion. We see
that for $t_L>>\gamma/\omega$ the dependence is
$$\approx 1/ (D\omega_0t_L)$$
\par The result for the arbitrary n,k is given by the derivative:
\begin{eqnarray}
<\vert a^n_k
(T)\vert^2>&=&H_n((\partial_{\lambda_1}+\partial_{\lambda_2})/2)
\times H_n((\partial_{\lambda_2}-\partial_{\lambda_1})/2)
\times
H_k((\partial_{\mu_2}-\partial{\mu_1})/2)\nonumber\\
&H_k&(\partial_{\mu_2}
+\partial_{\mu_1})/2) I(T,\mu_1=0,\mu_2=0,\lambda_1=0\lambda_2=0)
\label{dif}
\end{eqnarray}
Here the integral I is  \beq I=\int ds_0ds_1d\delta_0d\delta_1
\exp (iS+\mu_1s_0+\mu_2\delta_0
+\lambda_1\delta_0+\lambda_2\delta_1
-M\omega_0(\delta_1^2+\delta_0^2+s_1^2+s_0^2)/4 )/\sin (\omega_0
T) \label{i} \eeq The integral I is a gaussian integral that can
be easily taken: \beq I=\frac{1}{\sqrt H}\exp
(-Z/H-(\lambda_1^2+\mu_1^2)/M\omega_0))) \label{I} \eeq Here
\begin{eqnarray}
Z&=&(M\omega (1 +D(\omega T-\sin(2\omega
T)/2))(\mu_2^2+\lambda_2^2)\nonumber\\
&-&(M\omega(1+D(\omega T+\sin(2\omega
T)/2))(\mu_2^2+\lambda_2^2)\nonumber\\
&+&2iDM\omega\sin^2(\omega T)(\mu_1\mu_2+\lambda_1\lambda_2 )\nonumber\\
&-&2i(\lambda_2\mu_1+\lambda_1\mu_2)\sin (\omega T)M\omega(1 +D\omega T)+
2\mu_1\lambda_1M\omega(\cos (\omega T)+D\omega T+D\sin
(\omega T))\nonumber\\
&+&\mu_2\lambda_2M\omega (\cos (\omega T)+D\omega T-D\sin
(\omega
T)))/2H
\label{M}
\end{eqnarray}
\par The problem is of course 3Dimensional, and the explicit expression
for 3D coefficients are just a direct product of one-dimensional
coefficients in $x-y$ plane and in z direction
\beq
\vert a^{n_1n_2n_3}_{k_1k_2k_3}=\prod_{i=1,2}
a^{n_i}_{k_i}\delta^{n_3}_{k_3}
\label{pro}
\eeq
\par Here we took into account that for  sufficiently big $\gamma$ the
transverse
fields are strongly enhanced and we need only to take into account
the transverse transitions.
\par The evolution of the arbitrary initial distribution
(\ref{expansion}) can be calculated in the same way.
\section{Evolution of the Arbitrary  Wave Packets.}
\par In the previous section we considered the time evolution of the
arbitrary initial states. In this section we shall consider the
evolution of the small wave packet that has at t=0 the form in the
transverse plane: \beq \Psi
(y)=(2dM\omega_0/\pi)^{1/4}\exp(-dM\omega_0y^2) \label{small} \eeq
We first calculate explicitly the density matrix
$$\rho (x,y)=<\Psi (x,t)^*\Psi(y,t)>
$$ where the average means the averaging over all random field
configurations, and then use this expression to calculate the
probabilities. Substituting eq. (\ref{small}) into the equations
of the previous section, we get for the density matrix
\begin{eqnarray}
\rho(x,y,t)&=&\exp(-\delta^2(d/16+D^2d(\omega_0^2t^2-\gamma^2\sin^2
(\omega_0 t/\gamma)+(\omega t+\gamma \sin (2\omega_0
t/\gamma)/2.)D/4\nonumber\\
&+&d^2D(\omega_0 t -\sin (2\omega_0 t/\gamma)/2)-ds^2/16+2i\delta s
((d^2-1)\sin (2\omega_0 t/\gamma)/32\nonumber\\
&+&Nd\gamma \sin^2(\omega_0 t/\gamma)/4))/D(t))/\sqrt{D(t)}\label{density}
\end{eqnarray}
Here $D$ is given by eq. (\ref{D}) and the function $D(t)$ is given by
\beq D(t)=(d^2\sin (\omega_0
t/\gamma)^2+\cos (\omega_0 t/\gamma))^2/4.+Dd(\omega_0 t-\gamma
sin(2\omega_0 t/\gamma)/2) \label{diff3}\eeq
\par  We can determine
 density radius of the dipole/wave packet. The density radius
characterizes the decrease of $<\vert \Psi (x,t)^2\vert >$ as a
function of $x^2$. Note, that this is completely different
physical quantity than the effective radius of the dipole/wave
packet, that determines
 the effective cross-sections.
 \par The expansion of the
wave packet density is determined by $\rho (x,x)=<\Psi
(x,t)^*\Psi(x,t)>$ \beq x^2\approx D(t)/d=(d\sin (\omega_0
t/\gamma)^2+(1/d)\cos (\omega_0 t/\gamma))/4.+D(\omega_0 t-\gamma
sin(\omega_0 t2/\gamma)/2) \label{diff} \eeq Note that for the
small t$\ll \gamma/\omega$ we have \beq x^2(t)\approx
(1/d+\frac{3}{4}(d-1/d)\omega_0
t^2/\gamma^2+16D\omega_0^2t^3/\gamma^2)/4 \label{small1} \eeq
 We see
that there are three pieces in the latter equation: first, the
initial radius of the wave packet, second-the geometrical factor,
that is equal to zero when the particle is in the ground state,
and is bigger, the smaller is the wave packet radius (in
correspondence with the uncertainty principle). This factor
describes the wave-packet diffusion, that is present when there
are no interactions.
 Finally,
the third term is wave-packet radius independent and describes the
density expansion due to a strong field. We see that the latter for small
t goes as $t^3$, and for the case of the ground state $d=1$ this
is the only term that contributes to the expansion.
\par Before  the
time $t_L$ reaches $t_k\approx (d-1/d)/(\omega_0 D)$ the density
radius squared increases according to a square law, while for times $\geq
\gamma/\omega_0$ it increases according to the linear law.
\par We can determine, till what time scale we can consider the field
$\vec E$ as homogeneous. It is well known in the theory
of random systems \cite{B} that the field can be considered
homogeneous if  the inhomogeneity
scale in transverse direction is much bigger than time ( and z)
correlation scale.In
our case the latter scales  are $1/\kappa$ and $1/(\kappa\gamma )$
correspondingly. However, due to the longitudinal diffusion the
longitudinal-time correlation scale increases. Due to the fact
that the longitudinal electric field does not change strength due
to Lorentz boost, the expansion for the small packet in the
longitudinal direction is determined by the geometrical factor.
For the big packet with the radius of the ground state or bigger
the expansion rate will still be determined by the diffusion. The
longitudinal scale will reach the radius $1/\kappa$ at the minimum
of the times $\gamma^2/\sqrt{(d-1/d)\omega_0\kappa}$,
$\gamma^{4/3}/(\omega^2_0\kappa)^{1/3}$.
This means that for all realistic values of $d$ and for sufficiently
large $\gamma$, our results, derived for the
homogeneous fields, certainly hold up to the times of $\approx
\gamma/\omega$,
that is the characteristic time in the interaction of the dipole
and the target.
\par
\begin{figure}[htbp]
\centerline{\psfig{figure=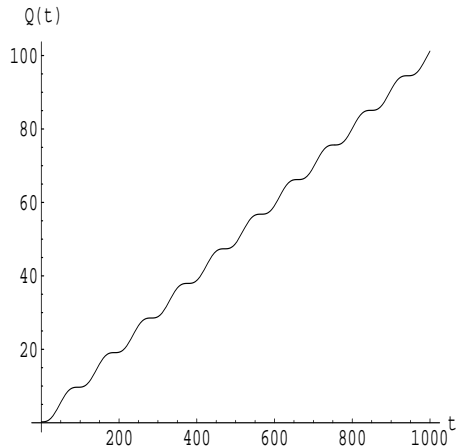,width=6cm,height=6cm,clip=}}
\caption{ Density radius of 1D oscillator in the random media as a
function of time for big time scales, if initially it was in the
ground state. Here and below $Q(t)=x^2(t)$} \label{ret8}
\end{figure}

\begin{figure}[htbp]
\centerline{\psfig{figure=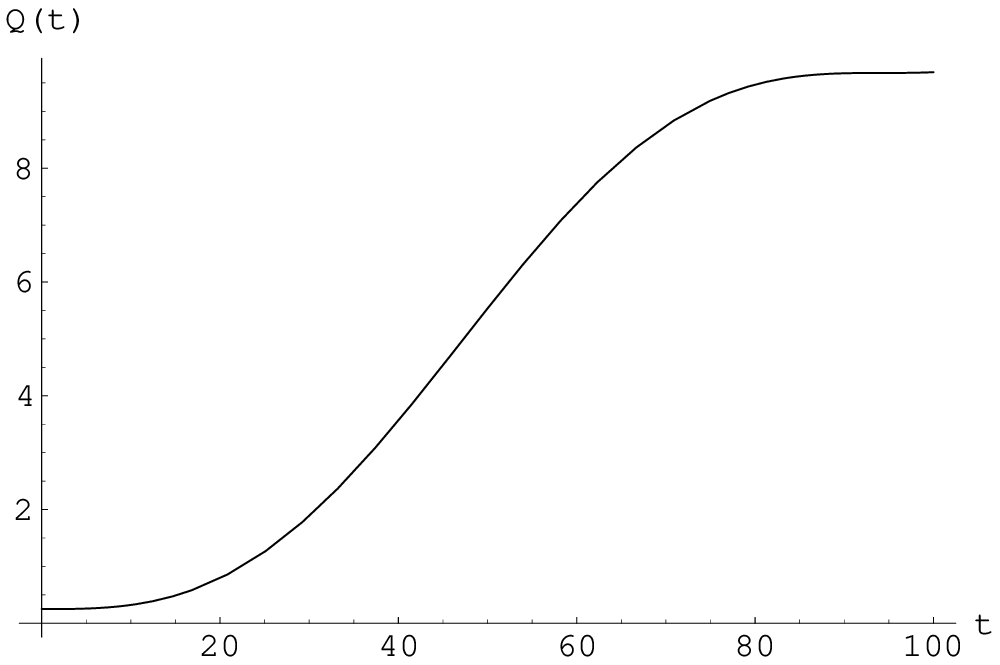,width=6cm,height=6cm,clip=}}
\caption{ Density radius of 1D oscillator in the random media as a
function of time for average time scales, if initially it was in
the ground state. }\label{ret9}
\end{figure}

\begin{figure}[htbp]
\centerline{\psfig{figure=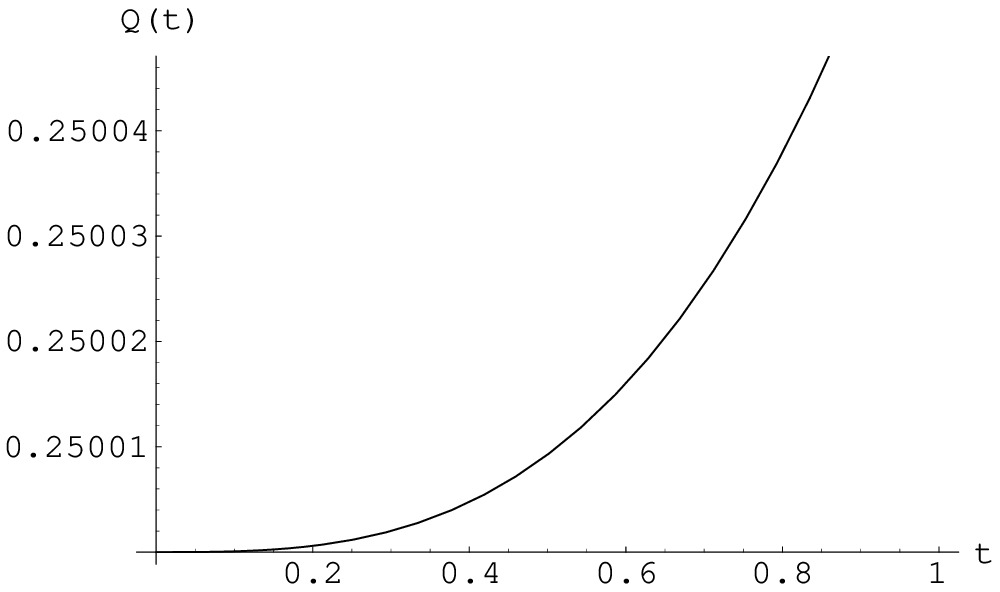,width=6cm,height=6cm,clip=}}
\caption{ Density radius of 1D oscillator in the random media as a
function of time for short time scales, if initially it was in the
ground state. }
\label{ret10}
\end{figure}

\begin{figure}[htbp]
\centerline{\psfig{figure=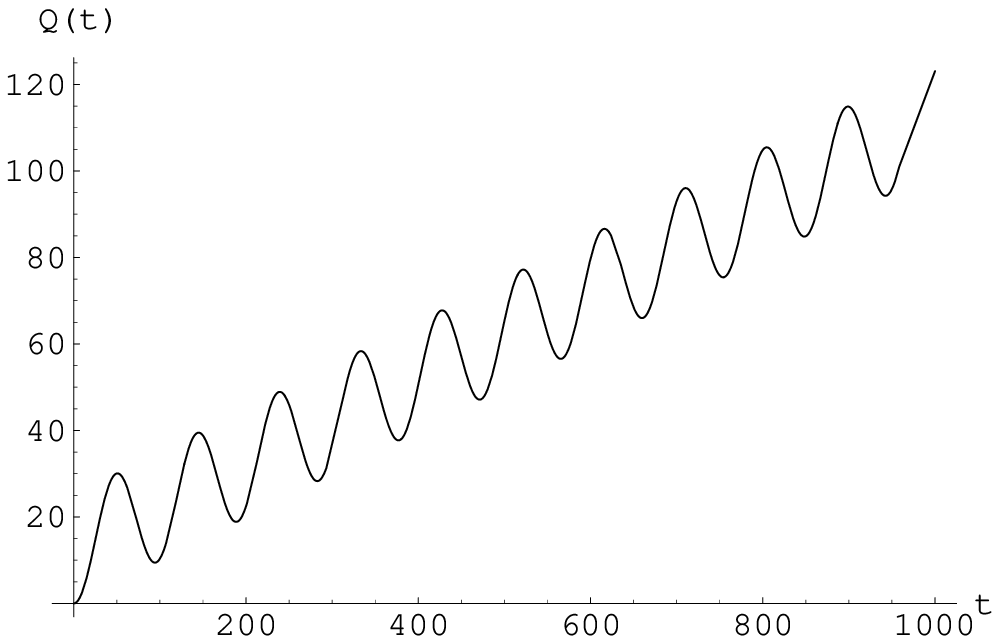,width=6cm,height=6cm,clip=}}
\caption{ Density radius of  oscillator in the random media as a
function of time for big time scales, if initially it was
 a small a=0.1 wave packet.}
\label{ret12}
\end{figure}

\begin{figure}[htbp]
\centerline{\psfig{figure=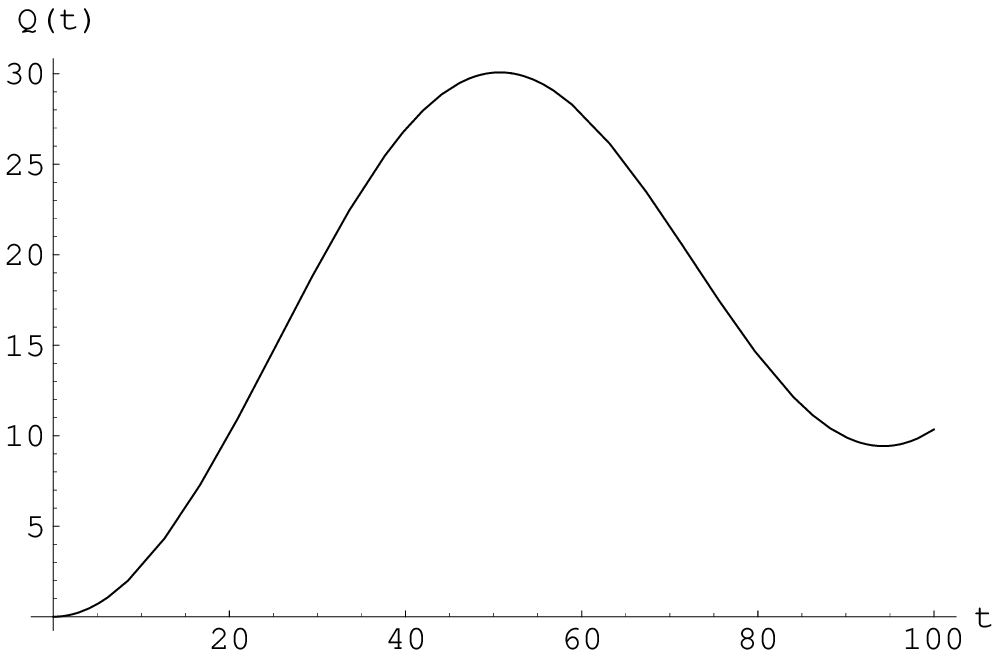,width=6cm,height=6cm,clip=}}
\caption{ Density radius of 1D oscillator in the random media as a
function of time for average time scales, if initially it was a
small a=0.1 wave packet. }
\label{ret13}
\end{figure}

\begin{figure}[htbp]
\centerline{\psfig{figure=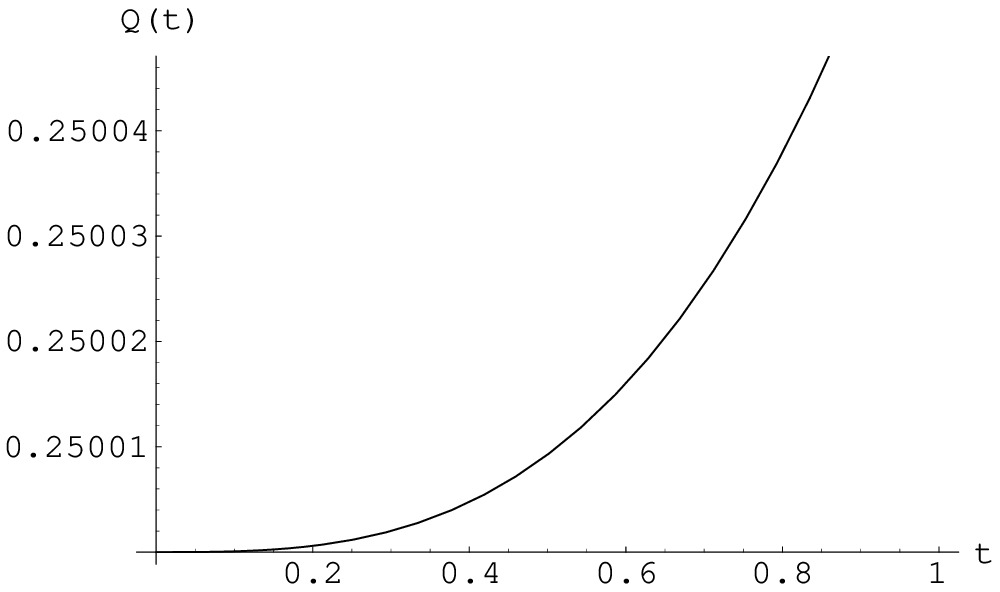,width=6cm,height=6cm,clip=}}
\caption{ Density radius of 1D oscillator in the random media as a
function of time for short time scales, if initially it was a
small a=0.1 wave packet.}
\label{ret14}
\end{figure}

 For the case of the oscillator the same result can be
obtained in the different way, using the Erenfest theorem. Indeed \beq
\frac{\partial^2 \vec
x}{dx^2}+\omega^2 \vec x=\vec E(t)/M \label{E} \eeq This operator
equation is easily solved: \beq \vec x(t)=A\cos (\omega
t_L/\gamma)+\int^t_0 ds\sin (\omega (t-s)) E(s) \vec /M \label{pitaron}
\eeq Then \beq <\rho^2
(t)>=(B^2\kappa\gamma))\int^t_0\int^t_0dsds'\sin (\omega
(t-s))\sin \omega (t-s')<E_z(s)E_z(s')> \label{averl} \eeq for
longitudinal diffusion, and \beq <x^2
(t)>=(B^2\gamma^2\kappa\gamma )\int^t_0\int^t_0dsds'\sin
(\omega (t-s))\sin \omega (t-s')<E_T(s)E_T(s')> \label{aver} \eeq
for transverse diffusion. In this way we obtain the law of
expansion: \beq <\rho^2 (t)>\approx B^2((2t_L-\gamma\sin(2\omega
t/\gamma)/(2\omega ) ))\times
\kappa^2\gamma^2/(\kappa^2\gamma^2+\omega^2)
 \label{diffusion}
\eeq We depicted the characteristic dependence of the square
radius of the oscillator in the initial ground state and in the
initial small wave packet state in figures
\ref{ret8},\ref{ret9},\ref{ret10} and
\ref{ret12},\ref{ret13},\ref{ret14} respectively.
 Note that in this derivation it is irrelevant if we consider
classical or quantum problem. The coincidence of the results of
two approaches is of course due to the Erenfest theorem.

\par
\begin{figure}[htbp]
\centerline{\psfig{figure=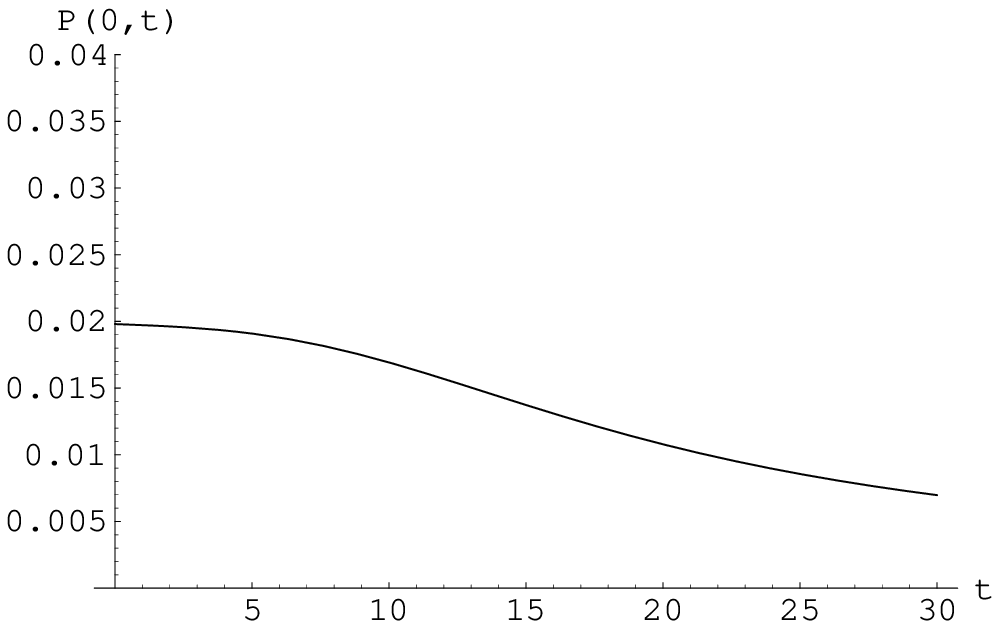,width=6cm,height=6cm,clip=}}
\caption{Probability of the oscillator that is a small wave packet
with radius a=0.1 (here and below in the ground state radius
units) to go from the ground to the tenth state as the function of
time. We use here and below: $\omega_0= 1$ GeV, $M= 1 $ GeV-the
reduced mass of the sample quarkonium, $\gamma=30$, and $\kappa
=0.12$ GeV; $B^2=0.012$ GeV$^4$. } \label{ret3}
\end{figure}

\begin{figure}[htbp]
\centerline{\psfig{figure=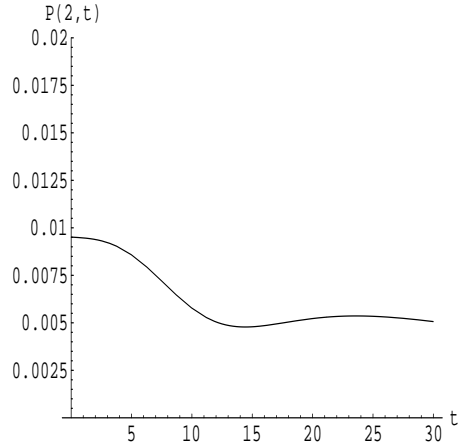,width=6cm,height=6cm,clip=}}
\caption{Probability of the  small oscillator wave packet with
radius a=0.1 to be in the second state at the  time t.}
\label{ret4}
\end{figure}

\begin{figure}[htbp]
\centerline{\psfig{figure=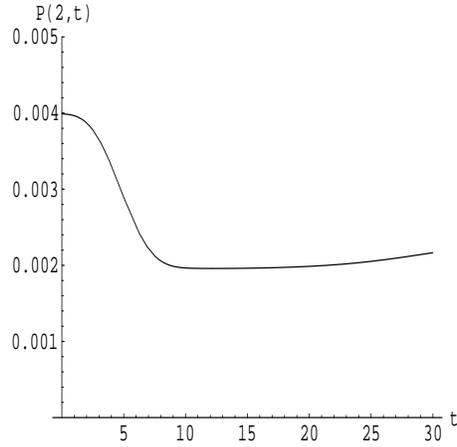,width=6cm,height=6cm,clip=}}
\caption{Probability of the small oscillator wave packet with
radius a=0.1 to be in the tenth state as the function of time.}
\label{ret5}
\end{figure}

 Once we know the laws of quantum expansion we can calculate the
probability distribution for the system to be in the n-th state.
The probability is given by \beq <\vert a_n(t)\vert
^2>=S(t)^{n/2}/R(t)^{(n+1)/2 }P_n(Q(t)/(2D(t)\sqrt{S(t)R(t)})
\label{prob} \eeq Here \begin{eqnarray}
Q(t)&=&D^2d(\omega^2t^2-\gamma^2\sin (\omega t/\gamma)^2)+D(\omega
t+\gamma \sin (2\omega t/\gamma)/2)/4+\nonumber\\
&N&d^2((\omega t -\sin (2\omega t/\gamma)/2)/4. \label{s}
\end{eqnarray} $P_n$ are the Legendre Polynomials, and D(t) is
given by eq.(\ref{diff3});
\begin{eqnarray} S(t)&=&-1/4-(D^2d^2(\omega^2
t^2-\sin^2 (\omega t/\gamma)\cos^2 (\omega
t/\gamma )+d^2/16\nonumber\\&+&Dd(\omega
t+\gamma\sin (4*\omega t/\gamma)/4)/4+Dd^3(\omega t-\sin (4\omega
t/\gamma)/4.)/4\nonumber\\& +&(d^2-1)^2\sin (2\omega
t/\gamma)^2/64)/(4D(t)^2))\nonumber\\
&(&d/8+D^2d(\omega_0^2t^2-\gamma^2\sin^2(\omega t/\gamma))+D(\omega
t+\gamma \sin (2\omega t/\gamma )/2)/4\nonumber\\
&D&d^2(\omega t-\sin (2\omega t/\gamma)/2)/4.)/D(t)
 \label{S} \end{eqnarray}

 \begin{eqnarray} R(t)&=&+1/4+(D^2d^2(\omega^2
t^2-\sin^2 (\omega t/\gamma)\cos^2 (\omega
t/\gamma )+d^2/16\nonumber\\&+&Dd(\omega
t+\gamma\sin (4*\omega t/\gamma)/4)/4+Dd^3(\omega t-\sin (4\omega
t/\gamma)/4.)/4\nonumber\\& +&(d^2-1)^2\sin (2\omega
t/\gamma)^2/64)/(4D(t)^2))\nonumber\\
&(&d/8+D^2d(\omega_0^2t^2-\gamma^2\sin^2(\omega t/\gamma))+D(\omega
t+\gamma \sin (2\omega t/\gamma )/2)/4\nonumber\\
&D&d^2(\omega t-\sin (2\omega t/\gamma)/2)/4.)/D(t)
 \label{R} \end{eqnarray} The sample
graphs are given below ( see figures
\ref{ret3},\ref{ret4},\ref{ret5}).
\section{The Ground State Evolution.}
\begin{figure}[htbp]
\centerline{\psfig{figure=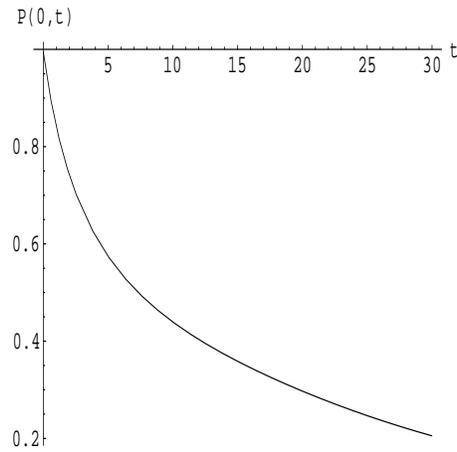,width=6cm,height=6cm,clip=}}
\caption{Probability of the oscillator to remain in the ground
state, if it enters the media in the ground state, as a function
of time. }\label{ret1}
\end{figure}

\begin{figure}[htbp]
\centerline{\psfig{figure=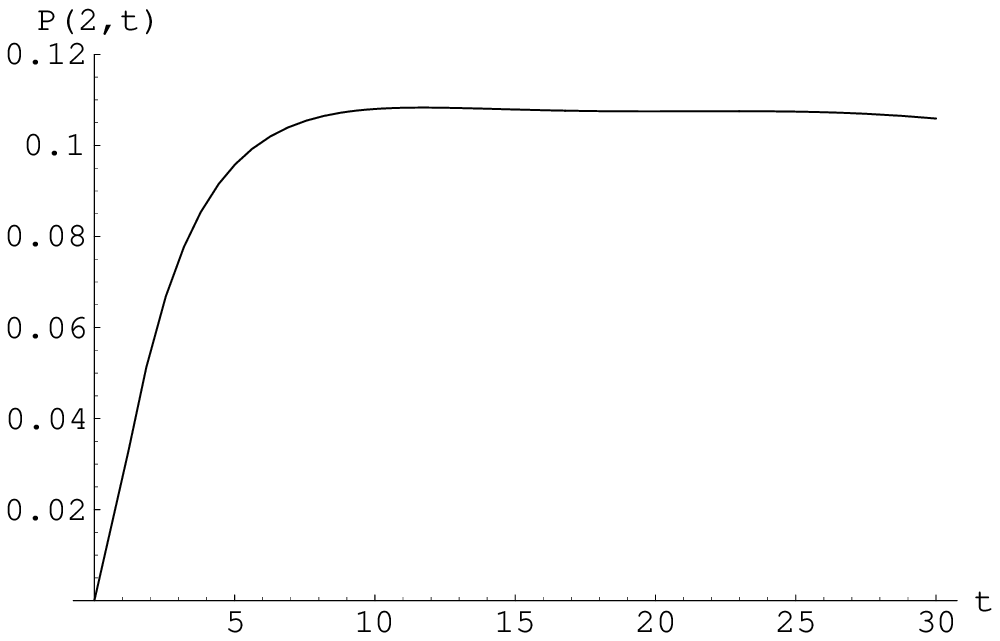,width=6cm,height=6cm,clip=}}
\caption{Probability of the oscillator to go from the ground to
the second state as the function of time.} \label{ret2}
\end{figure}

\begin{figure}[htbp]
\centerline{\psfig{figure=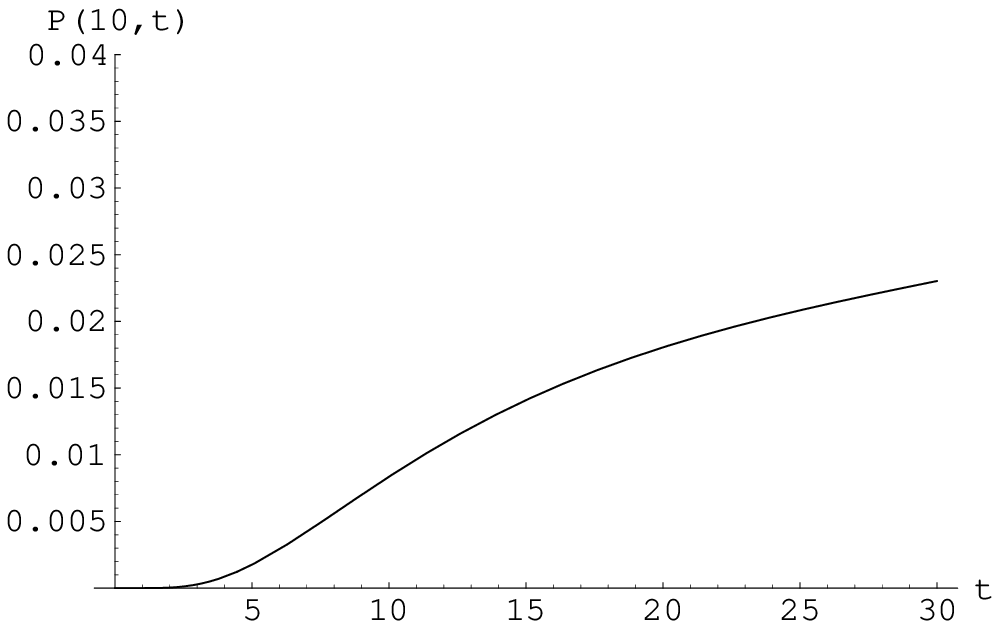,width=6cm,height=6cm,clip=}}
\caption{Probability of the oscillator to go from the ground to
the tenth state as the function of time.} \label{ret11}
\end{figure}

Consider now the oscillator entering the random media in the
ground state. Then we can put d=1 in the formulae of the preceding
chapter and all  results will be significantly
 simplified:
  \beq
R(t)=H(t)/(4D(t)),{\rm }\,\,\, S(t)=D^2(\omega_0^2t^2-\gamma^2\sin
(\omega_0 t/\gamma )^2/D(t)\label{simple}\eeq
 The
probability is given by \begin{eqnarray} <\vert a_n(t)\vert
^2>&=&((2D)^2(\omega_0^2t^2-\gamma^2\sin (\omega_0 t/\gamma
)^2)^{n/2}/(H(t))^{(n+1)/2
}\nonumber\\&\times&P_n(D\sqrt{(\omega_0^2t^2-\gamma^2\sin
(\omega_0 t/\gamma )^2/H(t)}) \label{prob1} \end{eqnarray}   The sample
graphs are given below (see figs. \ref{ret1},\ref{ret2},\ref{ret11}).
 \par These results can be applied to the simplest model of
charmonium
 moving through the Quark-Gluon plasma.
This picture of charmonium is of course  not realistic , since it
does not contain absorption \cite{FS,Kopeliovich,KM}, coulomb interactions
\cite{Susskind,Quigg},
color and radiation. It is given here for illustrative purposes only. The
more realistic picture of charmonium will be discussed elsewhere.
In our model the charmonium enters the media (say quark-gluon
plasma or nuclear matter) as the dipole in the ground state. The
charmonium is scattered by the random electric fields in the media
and is excited to the states with $N=2$ or higher states, where it
decays via $D\bar D$ pair decay mechanism. We assume for
simplicity that the decay probability is one, i.e. once the
charmonium crosses the $D\bar D$ threshold \cite{Susskind,Quigg},  the
total
probability that charmonium will still be charmonium and not decay
will be given by the sum \beq P(T)=\sum_{i=0,j=0,k=0}^{i+j+k\le
N_0}\vert a(0\rightarrow ijk)\vert^2 \label{char} \eeq
Here $N_0$ is the energy level corresponding to the $D\bar D$ threshold.
It is
enough to take into account only transitions in transverse directions,so
the sum is given by \beq P(t)=\sum_{p+s\le N_0} \vert a^p_0a^s_0
\vert^2 \vert \label{c} \eeq
\par
\begin{figure}[htbp]
\centerline{\psfig{figure=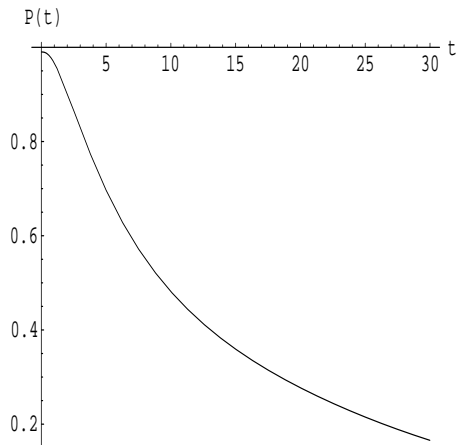,width=6cm,height=6cm,clip=}}
\caption{Probability of the charmonium entering the media in the
ground state to remain charmonium at  time t.} \label{ret6}
\end{figure}

 The result is depicted in figure \ref{ret6}.
 In the calculations we
have taken the realistic
 $\omega_0$ = 1 GeV.
For the realistic $\kappa$ we take 0.12 GeV, for $\gamma\approx
30$ (that corresponds to quick charmonium produced in the Fermilab
experiment), and for the field B we take the field in the QCD
vacuum \cite{SVZ} , that is say $0.11$ GeV$^2$, and for illustrative purposes
we took $N_0=2$.
\par Our results imply that this mechanism can make a significant
contribution in the dissociation of the rapid charmonium in the QCD media.
\par We also did not take into account that there exist, for the
small wave packet, backward transitions from higher levels to low
ones, that must be discarded in the realistic model (since
charmonium will dissociate). This question will be dealt
separately.
\section{Conclusion.}
\par We have studied the qualitative influence of the random media on
the relativistic propagation of the nonrelativistic dipoles and
wave packets with internal interaction ( see eqs.
(\ref{density}),(\ref{small1})).
 We showed that the media: 1) enhances the
expansion rate and leads to the new regime of the superdiffusion expansion
at the
rate $\rho^2\approx t^3$ for a ground state and for the small wave
packet for a sufficiently large time intervals if a field is
sufficiently large; 2) leads
to the diffusion over levels, leading to new types of diffractive events.
\par We have seen that the wave packet expansion rate depends on
the field strength and can be used to study the properties of the
media. Moreover, our results strongly suggest that the energy
losses and radiation of the dipole or wave packet propagating
through the random media will strongly depend on the media
(compare with the corresponding calculation for a color parton
\cite{D}), and can be used for the diagnostics of the quark-gluon
plasma. Our results also imply an additional mechanism of the
charmonium output suppression relative to the known ones
\cite{KM,Gerland}.
\par Our results imply the possibility of an additional mechanism of
dissociation of rapid charmonium by excitation to the higher levels due to
scattering on the random fields, and consequent dissociation into $D\bar
D$ pairs.
\par We have calculated the transition operator (\ref{G}),(\ref{finaction})) for the
oscillator, and calculated the time dependence of the
probabilities $P_n(t)$ of the oscillator occupying the n-th level
if it was in some initial state. The corresponding formula is
given by eqs.
(\ref{s}),(\ref{D}),(\ref{S}),(\ref{R}),(\ref{prob}). We used
these results to study the simple illustrative model of
charmonium, and were able to determine how many charmoniums will
stay after time t (or, equivalently, if a charmonium beam goes
through the shell of the width $L=v_0t$) (see eqs. (\ref{prob1}), (
\ref{char}).
\par The results obtained in this paper were derived for a particular
case of the moving oscillator. We expect that the same results
still hold, at least qualitatively, for more general system, such
as a Coulomb one.
 and. Indeed, the mechanism of superdiffusion expansion is the pick-up
by the system of the resonance frequencies from the media field
frequency spectrum. In the system other than oscillator, say in
the Coulomb system, it will pick the frequencies corresponding
to the distances between nearby levels, and consequently will
quickly expand, as in the simple oscillator case. It will be very
interesting and necessary for the study of realistic systems to
take into account absorption \cite{FS,Kopeliovich}.
\par We have calculated the effective action and transition operator
(\ref{final}) for the propagation of the general relativistic
dipole in the media. This action, obtained after integration out
of the electric field is quite general, and can be used for the
arbitrary bound systems. However, for the cases other than
harmonic oscillator it is hard to carry the exact calculations and
one may resort to numerical methods.
 \par It will be very interesting to use our method
to study a number of problems not considered here: the propagation
of positronium through amorphous media and it's ionization rate,
and the more realistic charmonium model that includes color and
absorption, as well as to the detailed analysis of the influence
of the nonhomogeneity. It will be also very interesting to carry
the calculation of the radiation loss and it's influence on the
dynamics of the dipole. At this point, due to the appearance of
the traces of the several color matrices, we shall expect real
technical difference between calculations in the usual plasma and
quark gluon plasma.

 \acknowledgements
The author is indebted to Prof. L. Frankfurt for numerous very
fruitful discussions and careful reading the paper. The author is
also thanks Dr. A. Kamenev for useful discussions.
\appendix
 \section{The Plasma Model and it's Statistical Description.}
\par The aim of this section is to connect the usual language of the
ref. \cite{DG} model with the language of the statistical
mechanics of the random media.
\par Let us recall briefly the model of (\cite{DG}).
The model describes the QCD media as the usual Debye plasma, with
a potential \beq V^a_i (\vec q)=g (T^a_i)_{cc'}\exp(-i\vec q\vec
x_i)/(\vec q^2+\mu^2) \label{pot} \eeq Here $\mu $ is the Debye
screening mass. The parton-parton interactions are proportional to
$tr (T^a_iT^b_j)$. The averaged potential vanishes everywhere
since $<V^a_i>\sim Tr T^a_i=0$, while the parton-parton
interactions are proportional to \beq
Tr(T^a_iT^b_j)=\delta_{ab}\delta_{ij}(d_i/d_A)C_{2i} \label{wang}
\eeq Here $d_i$ is the dimension of the SU(N) representation,
$C_{2i}$ is the second Casimir, ($(N^2-)/(2N)$ for quarks
($d_i=N$)  and N for gluons ($d_i=N^2-1$). Thus after averaging we
see that such model looks like multicomponent plasma, with a
particular interacting charge for each set of color partons, with
different types of partons not interacting in the leading order.
Thus, we expect that apart from some numerical factors, connected
with the Casimirs, the propagation of the wave packet in such a plasma
will be qualitatively the same as it's propagation through the
usual electromagnetic plasma described in the article. The true
color effects will appear only when we shall take into account
multiparticle correlations.
\par We now turn to the description of the charged plasma in the language
of the statistical mechanics (\cite{Feynman}). In order to do it,
we can just rewrite the statistical sum of the plasma as a
Euclidean Functional Integral. The choice of the action must be
such that it reproduces the correlation function
(\ref{correlation}). Clearly, the functional integral is Gaussian
and can be represented as \beq Z=\int dE\exp (-\int d^3qV
(q^2+\kappa^2)V) \label{connection} \eeq Moving into the
oscillator rest frame and taking into account that the time
derivative term will be dominant, we come to the action
(\ref{ac}).
 This description is just the usual language of the statistical mechanics
of the random media.
\section{Lorentz Transformations and Center of Mass Motion.}
\par Let us introduce the center of mass coordinates (for simplicity let us
assume that both particles have the same mass )
$v$ and the relative motion coordinate $u$:
\begin{eqnarray}
\vec v=(\vec r_1+\vec r_2)/2\\
\vec u=(\vec r_1-\vec r_2)/2\\
\label{coordinates}
\end{eqnarray}
We shall denote M the reduced mass of the oscillator in c.m.
reference frame. Then the Lagrangian can be written as a sum: \beq
L=\int ds (v) +L_{\rm n.r.}(\vec u)\label{sum} \eeq
 Indeed, the
full Lagrangian is \beq L=\int ds (r_1)+ds (r_2) -(V(\vec
r_1)-V(\vec r_2)+V(\vec u))dt \label{sto} \eeq Here \beq ds (\vec
r)=\sqrt{1-\Large (\frac{\partial \vec r}{\partial t}\Large)^2}dt
\label{par}\eeq We then write: \beq ds(\vec
r_1)=\sqrt{(1-(\frac{\partial (\vec u+\vec v)}{\partial t})^2)}dt
\label{w1} \eeq \beq ds(\vec r_1)=\sqrt{(1-(\frac{\partial (\vec
u-\vec v)}{\partial t})^2)}dt \label{w2} \eeq In these two
expressions we expand their r.h.s. in Taylor series in u around v.
Then we immediately obtain: \beq ds(r_1)+ds(r_2)=ds(v)+\int
dt({\dot{\vec v}^2)/(\sqrt{1-{\dot {\vec u}}^2}/2} =ds(v)+\int
dT({\dot{\vec v}}^2)/2 \label{sf} \eeq Here T is the oscillator
proper time. Adding the interaction term to the eq. (\ref{sf}) we
immediately obtain the decomposition (\ref{sum}).
\par Let us now discuss briefly the Lorentz transformation \cite{LLF}.
In the laboratory system we have only the electric field described by the
static potential
V. The Lorentz transformation says:
\beq
V=(V'-vA_3')\gamma\,\,\, A_3=(A'_3-vV')\gamma
\label{A}
\eeq
In the laboratory frame $A_3=0$, hence $A'_3=vV'$. Substituting the latter
formula in eq. \ref{A} we get
\beq
V=V'(1-v^2)\gamma=V'\gamma
\label{F}
\eeq
Finally we get
\beq
V'=V\gamma\,\,\, A_3'=vV'
\label{fa}
\eeq
Differentiating the latter equations in the moving reference frame we
obtain the transformation laws in the text.
\section{The Calculation of the Integral.}
\par Here for completeness we shall write the formulae for the double
integral used in the text:
\begin{eqnarray}
I&=&\int^{infty}_{-\infty}dx\int^\infty_{-\infty}dy\exp
(-Qx^2-Q^*y^2-2Pxy)
H_n(x)H_n(y)/(2^nn!)\nonumber\\
&(&Q+Q^*+P^2-1-\vert Q^2\vert)^{n/2}/(\vert Q^2\vert -P^2)^{(n+1)/2})\nonumber\\
&P_n&(-P/\sqrt{(\vert Q^2\vert -P^2)(Q+Q^*-\vert Q^2\vert
+P^2-1)}) \label{mainintegral}
\end{eqnarray}
The integral was calculated using ref. \cite{DP} The Legendre
Polynoms are defined as \cite{BE}:\beq
2^nn!P_n(x)=\frac{d^n}{dx^n}(x^2-1)^n \label{leg} \eeq



       
       

\end{document}